\documentclass{osa-article}

\journal{osajournal}

\articletype{Research Article}

\usepackage{lineno}

\begin{document}

\title{All-optical switching via coherent control of plasmon resonances}

\author{Rakesh Dhama,\authormark{1} Ali Panahpour,\authormark{2, 3} Tuomas Pihlava,\authormark{1} Dipa Ghindani,\authormark{1} and Humeyra Caglayan,\authormark{1*} }

\address{\authormark{1}Faculty of Engineering and Natural Sciences, Photonics, Tampere University, 33720 Tampere, Finland\\}
\address{\authormark{2}Javan Laser Company Ltd.  1464764511, Tehran, Iran}
\address{\authormark{3}JavanLaserCo@gmail.com}
\email{\authormark{*}humeyra.caglayan@tuni.fi} 



\begin{abstract*}
A novel ultrafast all-optical switching mechanism is demonstrated theoretically and experimentally based on a plasmonic analogue of the effect of \textit{Enhancement of Index of Refraction}(EIR) in quantum optics. In the quantum optical EIR the atomic systems are rendered by coherence and quantum interference to exhibit orders of magnitude higher index of refraction with vanishing or even negative absorption near their resonances. Similarly, in the plasmon induced EIR  a probe signal can experience positive, zero or negative extinction while strongly interacting with a metallic nanorod in a metamolecule which is coherently excited by a control beam. The same mechanism is observed in the collective response of a square array of such metamolecules in the form of a metasurface to modulate the amplitude of a signal by coherent control of absorption from positive to negative values without implementing gain materials or nonlinear processes. This novel approach can be used for challenging the control of light by light at the extreme levels of space, time and intensity by applying ultra-short pulses interacting with ultrafast surface plasmons or extremely low intensity pulses at the level of single photon to a nanoscale single plasmonic metamolecule. The scheme also introduces an effective tool for improving the modulation strength of optical modulators and switches through amplification of the input signal. 

\end{abstract*}

\section{Introduction}

Ultra-compact optical switches and modulators with ultrashort response time and minimal energy consumption are highly desirable in nanophotonics for development of efficient all-optical computing, processing devices and photonic networks\cite{chai2017ultrafast}. Conventional all-optical nanoswitches are mainly based on nonlinear effects such as Kerr, Raman, two photon absorption and frequency mixing. To improve the performance of nonlinear optical switches they can be equipped with plasmonic or resonant nanostructures to exploit their field enhancement and confinement features \cite{zhang2011multi,singh2020switching,lu2011ultrafast,ren2011nanostructured}. Other switching schemes involve hybridization of metamaterials by functional components such as semiconductors, carbon nanotubes, graphene, liquid crystals and phase-change materials \cite{dani2009subpicosecond,cho2009ultrafast,nikolaenko2010carbon,lim2018ultrafast,gholipour2013all,bergamini2021single}. However, the aforementioned switching mechanisms generally suffer from unsatisfactory response time and/or high power requirements. 

Linear interference phenomena are introduced as a promising approach for achieving ultra-fast all-optical switching\cite{fang2015controlling,plum2017controlling}  requiring arbitrarily low intensity optical beams down to the level of single photon regime\cite{roger2015coherent} . One realization is based on interference of two collinear, counter-propagating and phase-controlled coherent beams interacting with a plasmonic metasurface of deeply subwavelength thickness. The absorption of standing wave light is modulated by controlling the phase or amplitude of one of the counter-propagating lightwaves and consequently tuning the position of the nodes and antinodes on the metasurface \cite{fu2012all,zhang2012controlling}.   

As a more efficient approach, a class of linear optical nanoswitches and modulators are specifically designed to meet the requirements of the coherent perfect absorption (CPA) phenomenon\cite{wan2011time,noh2012perfect}   which is a time-reversed counterpart to laser emission and a generalization of the concept of critical coupling to an optical cavity  \cite{kang2014critical}. In a CPA system, complete absorption of electromagnetic radiation is achieved by controlling the interference of multiple incident waves. It can be realized in a variety of photonic structures, including planar and guided-mode structures, graphene-based systems, parity- and time-symmetric structures, epsilon-near-zero multi-layer films, quantum-mechanical systems and chiral metamaterials\cite{baranov2017coherent,kim2016general,ye2016coherent}. In a plasmonic nanostructure, CPA occurs when coherent light is completely absorbed and transferred to surface plasmons by exciting the nanostructure with the time-reversed mode of the corresponding surface plasmon laser or the so called SPASER \cite{noh2012perfect}. In experimental realizations of the effect, usually two symmetric plane waves incident on opposite sides of the system are completely absorbed, as a result of critical coupling into the dissipative degrees of freedom of the system. 

The classical counterparts of some quantum optical effects have also been widely implemented to develop the functionality of optical systems, metamaterials and switching devices \cite{kang2014critical,zhang2008plasmon,ling2018polarization} including the Fano resonance effect and electromagnetically induced transparency (EIT). In plasmonic systems, the plasmon induced transparency (PIT) creates a narrow transparent window within a broader absorption band of the system due to the destructive interference of super-radiant (radiative) and sub-radiant (dark) resonance modes of the plasmonic nanostructures. Typically, in an optical switch based on PIT, the transparency window of the medium is reversibly shifted in a wavelength range to provide the ON and OFF states of the switch through applying a light induced change in the refractive index of a nonlinear material around the plasmonic constituents\cite{bergamini2021single,fang2015controlling}. Recently, plasmonic analogue of a well-known effect in quantum optics named as \textit{enhancement of index of refraction} (EIR)\cite{scully1991enhancement} is theoretically reported in coupled plasmonic resonators, enabling the maximum susceptibility (strong electromagnetic response) with zero optical losses at resonance frequencies via coherent control of surface plasmons \cite{panahpour2019refraction}. Thus, apart form some newly proposed potential applications \cite{Gunay2020continuously,Yuce2020Ultra}, this approach has suggested a unique path to realize loss-compensated plasmonic devices operating at resonance frequencies through extraordinary enhancement of refractive index without using any gain media\cite{de2011dispersed,infusino2014loss,de2014double,dhama2016broadband} and nonlinear processes \cite{zhang2008plasmon,liu2009plasmonic}. 

In this work, we introduce and experimentally realize a novel ultrafast all-optical switching mechanism, based on this plasmonic analogue of the EIR effect. In contrast to the conventional methods of optical switching between zero and complete absorption limits of the system, our scheme is based on a linear phenomenon that enables switching of system absorption between positive and negative values in the absence of gain materials and nonlinear processes. This is achieved through coherent control of polarizability of nanoantennas by controlling the phase, polarization and amplitude of a control beam and results in unprecedented control in modulation strength of optical switches. The scheme is not restricted to counter-propagating light waves and the system can be illuminated by the signal and control beams from one side. 

We report on the design and fabrication of a particular plasmonic metasurface consisting of a square array of L-shaped meta-atom as an all-optical nano-switch whose response is decided via coherent control of surface plasmons. By proper tailoring of the phase, amplitude and polarization degrees of freedom of control beam, the dipolar polarizability of the nanoantennas excited by the signal beam can be linearly altered and it leads to control on overall signal transmission through the metasurface. Our approach shows how the absorption in the plasmon resonance band of metasurface can be controlled between positive and negative values as per demand using the properties of control beam as a tool. 

\section{Analytical model}
An analytical model is presented in this section to describe the proposed switching mechanism and estimate the performance of the nanoswitch, in the form of a plasmonic metasurface. To this end, a meta-atom is characterized consisting of two perpendicular nanoantennas in x and y directions of a Cartesian coordinate system illuminated by signal and control beams propagating along z direction. For simplicity of analytical calculations, we model the constituent nanoantennas of the meta-atom as identical spheroidal nanoparticles (NPs). The polarizability of each NP in the quasi-static approximation is given by:

\begin{figure}[h]
\centering\includegraphics[width=\linewidth]{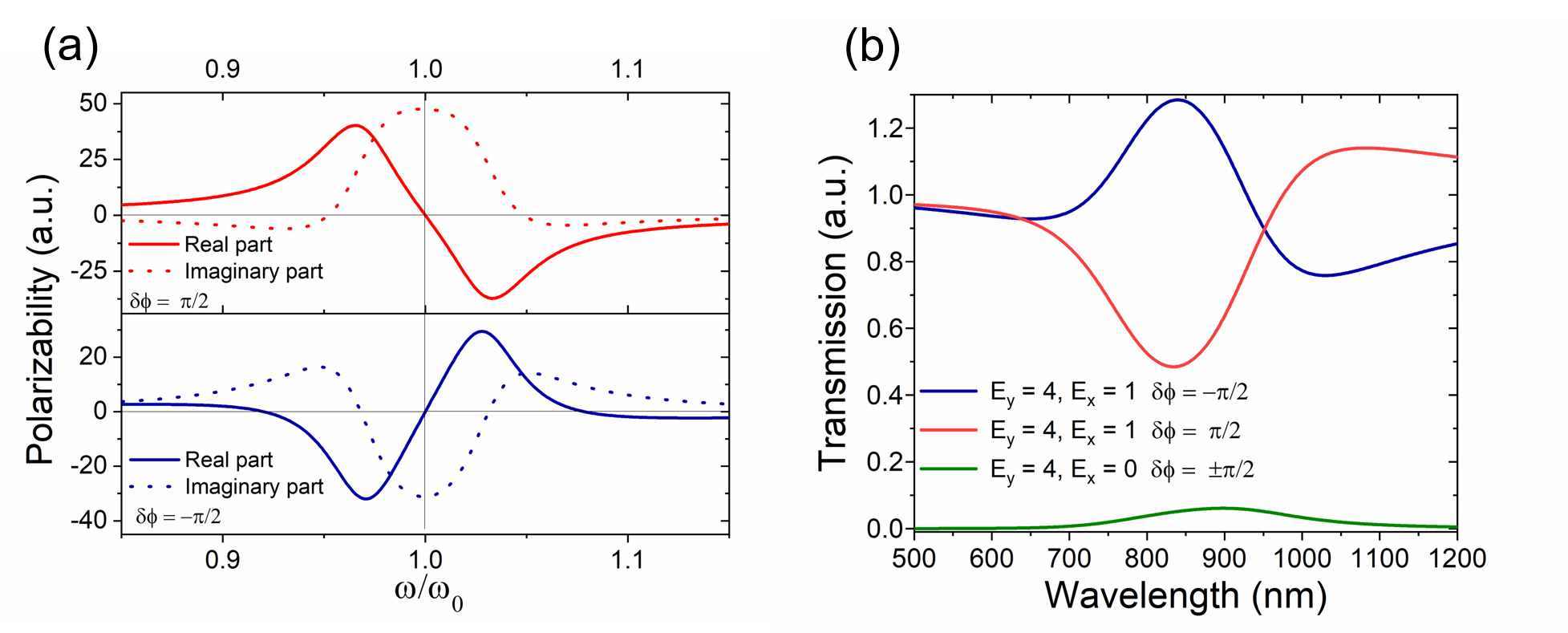}
\caption{(a) Real (solid lines) and imaginary (dotted lines) parts of polarizability of a spheroidal NP probed by a signal beam while nearly coupled to another orthogonally oriented identical NP which is excited by a phase correlated control beam. The control beam has either  \(\pi/2\) (upper curves) or  \(-\pi/2\) (lower curves) phase difference with respect to the signal beam. Geometrical parameters of \(a=75 nm\) and \(L=0.0284\) are used for the NPs with center to center distance of \(R=133 nm\). (b) The collective transmission response of the  meta-atoms corresponding to \(\pm\pi/2\) phase differences (blue and red curves) of the control beam. The green curve shows the leakage of control beam to the signal channel when signal is zero.}
 \label{fig:fig-1}
\end{figure}

\begin{equation}
\alpha^{-1}=\alpha_0^{-1}-i\frac{2k^3}{3}
\end{equation}

in which   \(-2ik^3/3\)  is the radiation correction term for the NP’s electrostatic polarizability \(\alpha_0\), and \(k\) is wavenumber. The electrostatic polarizability for each NP in free space is given by: 
\begin{equation}
\alpha_0^{-1}=\frac{1}{\nu}\left(L+\frac{1}{\epsilon-1}\right)
\end{equation}
where, \(\nu\)  and \(\epsilon(\omega)\)  are the volume and frequency dependent dielectric function of NPs respectively, and \(L=\frac{1-e^2}{e^2}(-1+\frac{1}{2e}ln\frac{1+e}{1-e})\) is an electrostatic geometrical factor in terms of eccentricity \(e=\sqrt{1-b^2/a^2}\) of the spheroids with \(a\)  and \(b\)  as their respective longer and shorter semi-axes. Using the Drude formula, \(\epsilon=1-\omega_p^2/\omega(\omega+i\gamma)\) for dielectric function of the NPs and normalizing the angular frequency and relaxation constant to plasma frequency as \(\tilde\omega=\omega/\omega_p\)  and \(\tilde\gamma=\gamma/\omega_p\) , from relations (1) and (2) we obtain:
\begin{equation}
\alpha^{-1}=\frac{1}{\nu}(L-\tilde{\omega}^2-i\tilde{\gamma}\tilde{\omega})-i\frac{2k^3}{3}
\end{equation}

The structure is illuminated by a x-polarized "signal" beam and a y-polarized "control" beam of the same frequency but larger amplitude. Assuming geometrical parameters of \(a=75 nm\) and \(L=0.0284\) for the NPs and center to center distance of \(R=133 nm\), the real and imaginary parts of polarizability of the nanoantenna along x-direction, probed by the x-polarized signal are shown in Fig. \ref{fig:fig-1}(a) as a function of normalized frequency. The curves correspond to the conditions when the control beam takes phase difference values of \(-\pi/2\) and \(\pi/2\) with respect to the signal beam. The curves show that the extinction of the probe beam (which is proportional to the imaginary part of polarizability) can take positive to negative values depending on the phase of control beam. It is also interesting that the extinction is zero at frequencies near extreme points of real polarizability. This particular profile of polarizability corresponds to the EIR phenomenon in contrast to the EIT effect, where the medium shows minimal extinction at the zero polarizability point. 

To study the collective response of the meta-atoms as an optical nanoswitch we consider a metasurface consisting of a square array of such meta-atoms. In this two-dimensional array of  meta-atoms, the induced dipole moment of each nanoantenna in x or y directions is a result of contributions from external field components along the respective x and y directions (\(E_{x,y}\)), as well as the electric field due to the closely coupled orthogonal nanoantenna (with coupling coefficient \(C\) ), and finally the fields due to the collective effect of other parallel (\(S_\parallel p_{x,y}\) ) and perpendicular ( \(S_\perp p_{x,y}\)) nanoantennas in terms of the corresponding lattice sums \(S_\parallel\)  and \(S_\perp\) :
\begin{equation}
p_x=\alpha(E_x+Cp_y+S_\parallel p_x+S_\perp p_y)
\end{equation}
\begin{equation}
p_y=\alpha(E_y+Cp_x+S_\parallel p_y+S_\perp p_x)
\end{equation}
which can be represented in the matrix form as:
\begin{equation}
 \left(\begin{array}{c}
p_x\\
p_y\end{array}\right)=\left(\begin{array}{cc}
\alpha_{xx} & \alpha_{xy} \\
\alpha_{xy}  & \alpha_{yy} \end{array}\right)\left(\begin{array}{c}
E_x\\
E_y\end{array}\right)
\end{equation}
where:
\begin{equation}
\hat{\alpha}=\left(\begin{array}{cc}
\alpha_{xx} & \alpha_{xy} \\
\alpha_{xy}  & \alpha_{yy} \end{array}\right)=\frac{\alpha^\prime }{1- {C^\prime} ^2{\alpha^\prime}^2}\left(\begin{array}{cc}
1 & C^\prime\alpha^\prime\\
C^\prime\alpha^\prime & 1\end{array}\right)
\end{equation}

here,  \(\alpha^\prime=\alpha/(1-\alpha S_\parallel)\), \(C^\prime=C+S_\perp\) and the coupling coefficient \(C=3/8\pi\epsilon_0R^3\) represents the near field interaction of coupled nanoantennas with center to center distance of \(R\). We note that the polarizability of each NP is affected by the radiated fields of all other parallel NPs, while the field due to the perpendicular NPs just modifies the coupling constant \(C\). It can be shown that the dipolar fields due to the perpendicular NPs symmetrically located on either side of each nanoantenna neutralize the effect of each other and therefore for a metasurface consisting a large number of meta-atoms the contribution of \(S_\perp\)can be neglected. For normal incidence on the metasurface with the dense array of meta-atoms ( \(kd<1\)), the interaction constant \(S_\parallel\), representative of the effect of all dipoles parallel to a specific nanoantenna is given by \cite{viitanen2002analytical}:

\begin{equation}
S_\parallel\approx-i\frac{\omega}{A}\frac{\eta}{4}\left(1-\frac{1}{ikD}\right)e^{-ik{D}}
\end{equation}
Here,  \(A=d^2\) is the unit cell area, \(\eta=\sqrt{\mu_0/\epsilon_0}\)  , \(k=\omega\sqrt{\mu_0\epsilon_0}\)  and \(D=d/1.438\) . 

Finally a transmission matrix of the form: 
\begin{equation}
\left(\begin{array}{c}
E_{tx}\\
E_{ty}\end{array}\right)=\left(\begin{array}{cc}
\ T_{xx} & \ T_{xy} \\
\ T_{xy}  & \ T_{yy} \end{array}\right)\left(\begin{array}{c}
E_{ix}\\
E_{iy}\end{array}\right)
\end{equation}
can be defined with the matrix components \cite{alu2011optical}:
\begin{equation}
T_{xx}=T_{yy}=1+\frac{i k\alpha_{xx}}{2\epsilon_0 d^2}
\end{equation}
\begin{equation}
T_{xy}=-\frac{i k\alpha_{xy}}{2\epsilon_0 d^2}
\end{equation}

to calculate the transmission through the metasurface. 

Assuming the same geometrical parameters as in Fig. \ref{fig:fig-1} (a) and unit cell dimension of \(d=320 nm\), we obtain the signal transmission curves shown in Fig. \ref{fig:fig-1}(b) for \(E_{signal}=1(a. u.)\) and control beams of amplitudes \(E_{control}=4e^{\pm i\pi/2}\). For the sake of simplicity we ignored the effect of substrate in our analytical calculations. 

We note that the maximum transmission can be higher than \%100 which is representative of signal amplification. The amplification strength depends on the phase and amplitude of the control beam. This amplification feature provides the capability of improving the modulation strength of the switch. However, the control beam amplitude cannot be arbitrarily increased. It is limited by crosstalk which is another measure of performance of a switch and is a representative of optical leakage from control channel to signal channel. In Fig. 1(b) the transmission of control beam through the signal channel is also plotted (the green curve) when there is no input signal. This leakage should be reduced to a small fraction of the signal strength.

\section{Experimental section}
\begin{figure}[h!]
\centering\includegraphics[width=\linewidth]{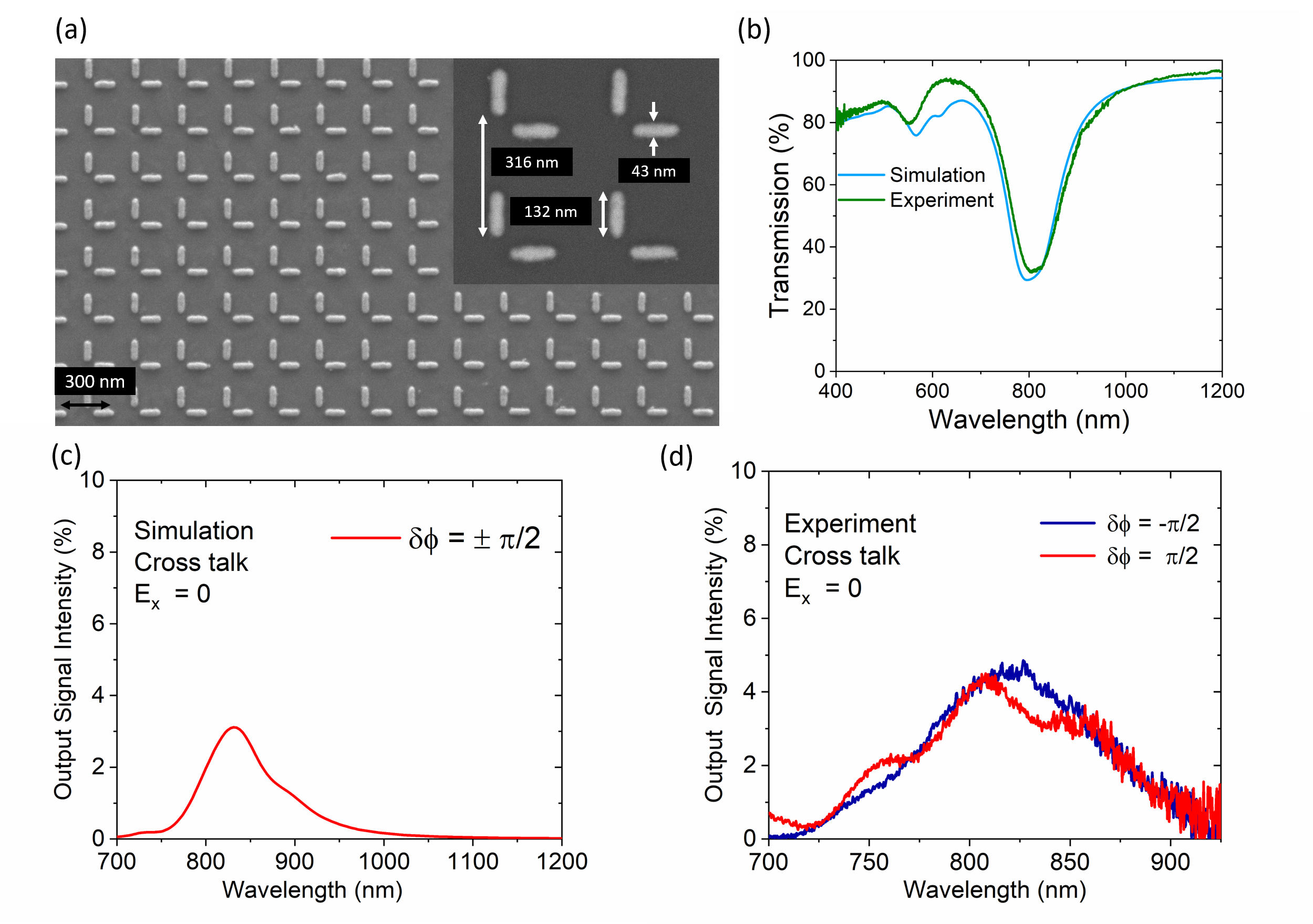}
\caption{(a) SEM image of the fabricated L-shaped metasurface, the inset of figure indicates the dimensions of nanorods. (b) Experimental and calculated transmission spectra of the metasurface without using any phase delay component and polarizer. (c,d) Investigation of cross-talk to measure the fraction of control beam, contributing in the output signal intensity.}
 \label{fig:fig-2}
\end{figure}

This plasmonic metasurface has been realized by fabricating two identical gold nanorods of dimensions 130 $\times$ 40 $\times$ 30 nm$\textsuperscript{3}$ (\textit{length $\times$ width $\times$ height}) as L-shaped meta-atom with a period of 320 nm on a fused silica glass substrate by electron beam lithography process. Scanning electron microscopy (SEM) images for the fabricated metasurface is shown in Fig. \ref{fig:fig-2}(a) visualizing the pattern of two closely placed orthogonal nanorods. The inset of Fig. \ref{fig:fig-2}  (a) indicates the dimensions of the nanorods identical to the values used in the simulations.

Fig. \ref{fig:fig-2} (b) reports intrinsic response when no phase delay component and  polarizer in optical system. Experimental (solid green line) and simulated (solid black line) transmission spectra of the metasurface under the illumination of broadband, incoherent and unpolarized light source shows the induced plasmon resonance around 800 nm. 
Slight red shift in the experimental transmission spectrum with respect to the simulated one can be attributed to minor fabrication imperfections, particularly the edges of nanorods.       

Before the investigation of the switching performance, we have confirmed theoretically as well as experimentally the cross-talk contribution in the output signal (x-component) intensity by illuminating the meta-structure with an excitation beam containing null signal beam (E\textsubscript{x} = 0) and a strong control beam (E\textsubscript{y} = 4).  The cross talk is an unwanted leakage of control beam in the output signal intensity which ideally should be zero when the input  signal beam is absent. Minimized value of cross talk is the measure of purity of the signal and an important factor for an efficient optical switch. Moreover, nominal cross-talk is the  key for the excellent performance of the devices such all-optical logic gates, optical circuits and quantum computing which can also leads to the development optical computers operating in ultra high speed network\cite{sasikala2018all}. We have obtained 5$\%$ cross-talk as demonstrated in  Fig. \ref{fig:fig-2} (c) and (d) which is quite lower in comparison to other cases reported in the literature \cite{davis2014all}.

\begin{figure}[h!]
\centering\includegraphics[width=\linewidth]{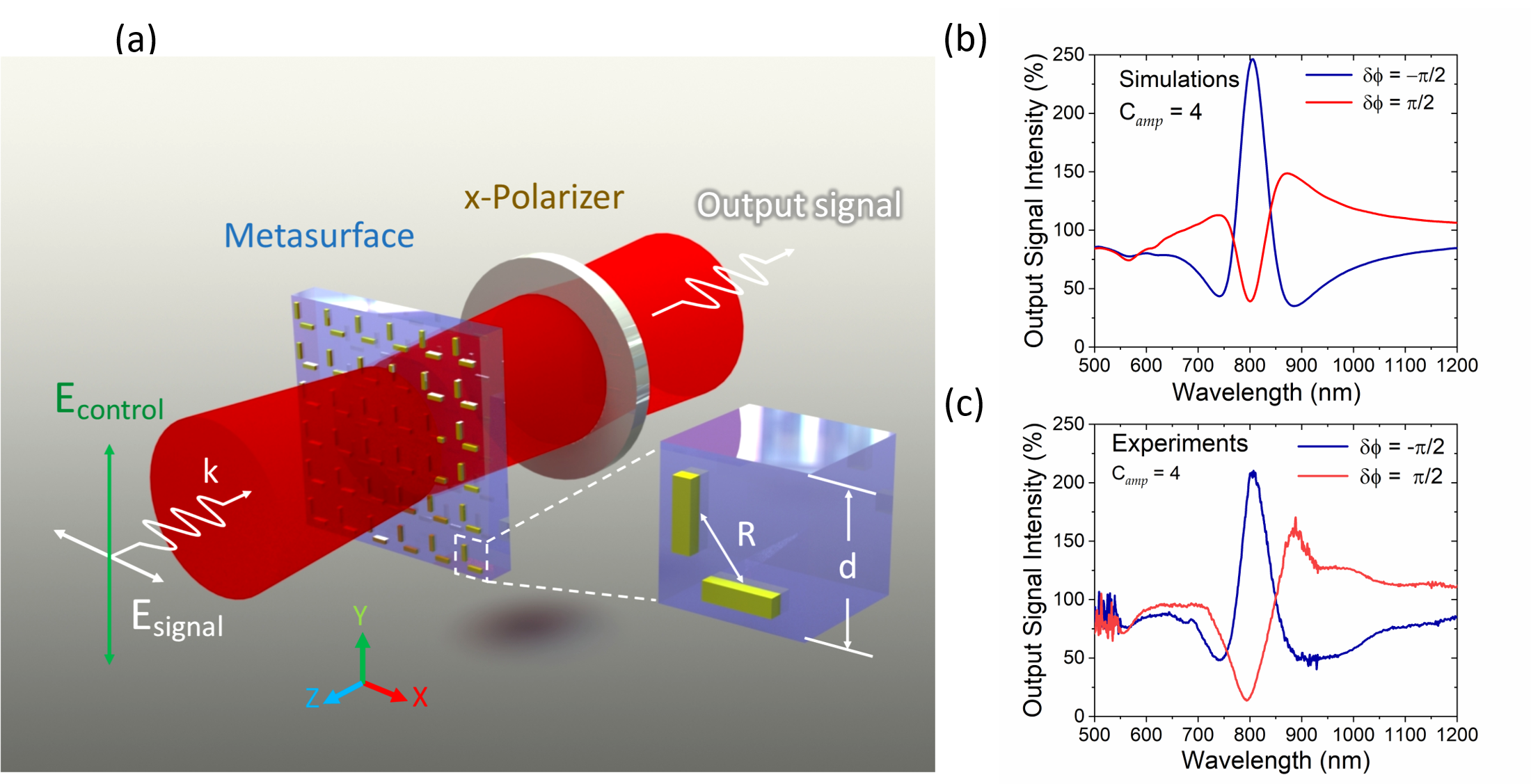}
\caption{All-optical nanoswitch based on coherent control of surface plasmons: (a) Schematic illustration for the modulated output signal intensity of device under the excitation of phase and polarization induced incoherent and broadband light beam. Inset of figure represents the schematic of a meta-atom from L-shaped metasurface where horizontally polarized light (\textit{x-pol}) represents the signal beam, while  vertically polarized light (\textit{y-pol}) acts as the control beam.(b,c) Simulated and experimental modulated output signal intensity at PC = 4  as the function of phase difference of  $\pi/2$ between control and signal beam, exhibiting the strong switching effect.}
 \label{fig:fig-3}
\end{figure}

To demonstrate the potential of this metasurface as an all-optical nanoswitch, Figure \ref{fig:fig-3} (a) represents the schematic for the used optical configuration. A broadband incoherent light beam travels through a linear polarizer and a phase delay component (quarter wave plate) to transform into phase and polarization controlled beam and excites the metasurface. The incident beam on each meta-atom of the metasurface can be described as two input beams which are distinguished by their orthogonal polarizations. Horizontally polarized beam along x-direction is entitled as ‘signal beam’, while vertically  polarized beam along y-direction is called ‘control beam’ as presented in the inset of  Fig. \ref{fig:fig-3} (a) of a single meta-atom. Another linear polarizer placed after the metasurface in the transmitted path ensures removal of any unwanted control beam component in the output and only allows the  output signal beam to direct to the detector.
The ratio of amplitudes of control beam to signal beam is defined as a Polarization Control (PC) parameter which also refers to specific polarizer angle.

The output signal intensity (x-polarized intensity) was numerically simulated when a polarized control beam with amplitude 4 times to signal beam (E\textsubscript{y}/E\textsubscript{x}= 4) along with a phase difference ($\delta\phi$) of $\pi/2$ excites the metasurface as shown in Figure \ref{fig:fig-3}(a). Remarkable enhancement in signal output intensity has been theoretically calculated with phase difference $\delta\phi$ = -$\pi/2$  (Fig. \ref{fig:fig-3}(b)) with respect to $\delta\phi$ = $\pi/2$, at the plasmon resonances of the metasurface. This predicts the strong numerically calculated modulation in signal beam due to the change in phase and polarization of control beam.  
 
Experimental realization of this switching effect is demonstrated in Fig. \ref{fig:fig-3} (c) when  an excitation beam is passed through a polarizer set at an angle  $\theta$ = $\tan\textsuperscript{-1}(Ey/Ex)$ ($\theta$ = 76$^{\circ}$  for PC = 4) and through a  quarter waveplate before illuminating the device and then signal beam is transmitted through the optical configuration as shown in schematic of Fig. \ref{fig:fig-3} (a). The experimental procedure to ensure the electric field ratio of control beam to signal one (PC) to be a certain value and alignment of linear polarizers used in excitation path and in transmission path have been discussed in detail in supplementary section. Figure \ref{fig:fig-3} (c) clearly demonstrates the strong switching effect (blue curve over red one) by inducing the giant transmission enhancement in the whole plasmon resonances band supported by L- shaped meta-atoms when phase of control beam lags behind the signal beam with phase difference $\pi/2$ ($\delta\phi$ = -$\pi/2$) and enables modulation in output signal intensity around 200 $\%$.  We note that coherent laser sources are not required to achieve coherent control of surface plasmons. Modulation strictly depends on PC and phase difference between control and signal beams which must be coherent with respect to each other. 


The extraordinary enhancement in output signal intensity (horizontally polarized light) is attributed to  electromagnetically induced negative absorption in the horizontal nanoantenna via coherent control of surface plasmon resonance. In order to understand the physical mechanism involved  behind this outstanding effect, we must revisit a classical analogue of the EIR effect in quantum optics, which states that atoms prepared in coherent superposition state can induce large resonant index of refraction with vanishing absorption via quantum interference between an excited state and coherently prepared ground state doublet\cite{scully1991enhancement}. Recently, the plasmon-analogue of the EIR effect is theoretically shown in two perpendicular cigar-shaped silver spheroidal nanoparticles when excited by two orthogonally polarized field with the phase difference $\delta\phi$ = $\pi/2$ and reported lossless resonance of localized surface plasmon resonance (LSPR) of a nanoantenna. When both nanoantennas of L-shaped meta-atom are coupled under the illumination of incident light beam and fulfill the mutual coherence condition then surface plasmons excited in vertical nanoantenna (control) induce an extraordinary modification in the resonance profile of the horizontal nanoantenna  probed by the x-polarized  signal beam. This modification of spectral profile of LSPR and induced zero or negative absorption is through mutual coupling of the nanoparticles and canalization of energy from vertically to horizontally oriented nanoantenna due to phase shift between signal and control beams. Thus, this energy exchange enhances the output signal intensity and enables the all-optical switching effect.

\begin{figure}[h!]
\centering\includegraphics[width=\linewidth]{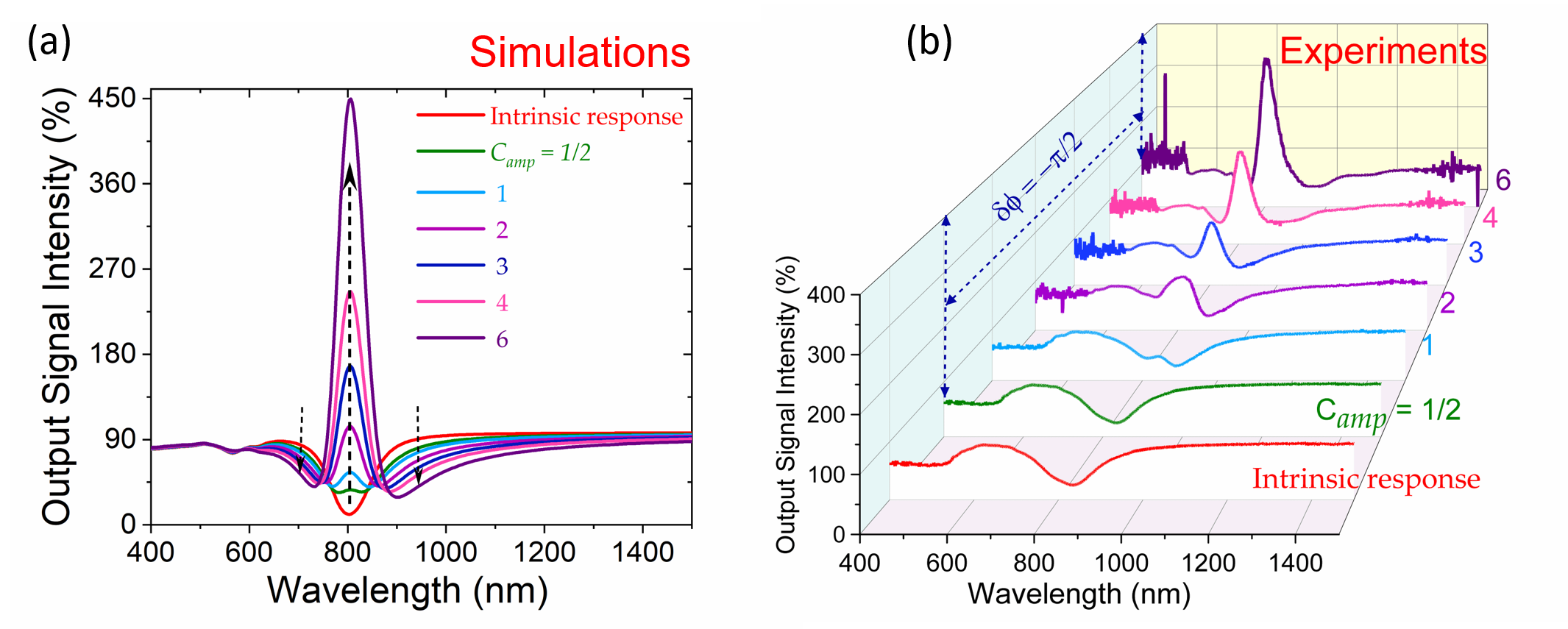}
\caption{Broadband transmission enhancement in the plasmon resonances band  of L-shaped metasurface: (a) Simulated output signal intensity as the function of Polarization Control (PC) with phase difference of  $\delta\phi$ = -$\pi/2$ (b) Transmitted output signal intensity curves as the function of Polarization Control (PC) with phase difference of $\delta\phi$ = -$\pi/2$ in 2D representation. (c) Experimental output signal intensity as the function of Polarization Control (PC) with phase difference of  $\delta\phi$ = -$\pi/2$.}
 \label{fig:fig-4}
\end{figure}

Figure \ref{fig:fig-4} shows how polarization control can determine the transparency in the absorptive plasmon resonances band of the L-shaped metasurface. Simulated output signal intensity curves clearly show a trend in the  transmission enhancement as shown in Fig \ref{fig:fig-4} (a)  with the increase in the values of PC in comparison to intrinsic plasmon band (red solid line) of this meta-device. Noted, intrinsic response of the device means when there is neither phase delay component nor polarizer in the optical configuration. Furthermore, we have plotted the similar simulated curves in 2D form as shown in Fig. \ref{fig:fig-4} (b) to clearly demonstrate that the transmission enhancement in the plasmon band of the metasurface (upward dotted black arrow) emerges at the cost of enhanced absorption in both arms (downward dotted black arrow) of output signal intensity curves. This behaviour can be explained with the causal nature of the response of metasurface via Kramers–Kronig (KK) dispersion relations and is also in analogy with a gain-plasmonic system where the reduction of absorption in selective region of plasmon band leads to an increase in the absorption in its neighbouring wavelength band  due to the decrease in imaginary part of permittivity via exciton-plasmon coupling effect\cite{de2012gain}. In a similar way, when the imaginary part of permittivity is reduced due to the EIR effect, transmission enhancement in the plasmon band of metasurface leads to decrease in the transmission on the both sides of transmission curves of output signal intensity. Fig. \ref{fig:fig-4} (c) demonstrates the experimentally realized curves in 3D waterfall form  which are in excellent agreement with the simulated results as shown in Fig \ref{fig:fig-4} (a).     

This index enhancement effect via coherent control of surface plasmons also shows a technique to the control absorption (optical losses) in the plasmon resonances band without involving any gain media. Our experimental and simulated results clearly show that this technique creates the transparency in complete plasmon band unlike a narrow transparency window in plasmon induced transparency. This approach can be implemented across the entire visible and near-infrared region to minimize the optical losses  using different metal nanostructures and specifically engineered design. On the other hand, controlling the absorption through such metasurfaces can realize the novel devices such as variable attenuators, coherence filters and coherent modulators, operating at arbitrarily low power levels.




\section{Conclusion}
In conclusion we have introduced a novel all-optical switching mechanism based on a plasmonic counterpart of the quantum optical effect of EIR. This is realized by utilizing phase correlated pump and probe optical beams of orthogonal polarizations interacting with a metasurface consisting of L-shaped meta-atoms. In this switching method, the absorption and dispersion of LSPR of the constituent nanoantennas are coherently and linearly controlled so that extinction of the plasmonic system probed by the input signal in one polarization direction can take positive or negative values depending on the phase of the orthogonally polarized control beam and thus enabling all-optical switching of the metasurface transparency from low levels to much higher than \%100. This novel switching scheme provides significant improvement in modulation strength while having reasonable low levels of crosstalk. 
The switching mechanism is described using a simple analytical model and also FDTD simulations. Based on the theoretical and simulation results the plasmonic metasurface nanoswitch was fabricated and characterised in the NIR spectral region. The results of the analytical calculation, simulation and experiment are in good agreement.


\begin{backmatter}
\bmsection{Funding}
We acknowledge the financial support of the European Research Council (Starting Grant project aQUARiUM; Agreement No. 802986), Academy of Finland Flagship Programme, (PREIN), (320165) and Marie Skłodowska-Curie MULTIPLY Project (Agreement No. 713694).

\bmsection{Acknowledgments}
R.D. and A.P. contributed equally to the work. The authors thank Subhajit Bej for the fabrication support.

\bmsection{Disclosures}

\noindent The authors declare no conflicts of interest.

\bmsection{Data Availability Statement}
The data that support the findings of this study are available from the corresponding author upon reasonable request.

\bmsection{Supplemental document}
See Supplement Material for supporting content. 

\end{backmatter}


\bibliography{sample}

\begin{thebibliography}{10}
\newcommand{\enquote}[1]{``#1''}

\bibitem{chai2017ultrafast}
Z.~Chai, X.~Hu, F.~Wang, X.~Niu, J.~Xie, and Q.~Gong, \enquote{Ultrafast
  all-optical switching,} {\protect\JournalTitle{Advanced Optical Materials}}
  \textbf{5}, 1600665 (2017).

\bibitem{zhang2011multi}
Y.~Zhang, X.~Hu, H.~Yang, and Q.~Gong, \enquote{Multi-component nanocomposite
  for all-optical switching applications,} {\protect\JournalTitle{Applied
  Physics Letters}} \textbf{99}, 141113 (2011).

\bibitem{singh2020switching}
M.~R. Singh and S.~Yastrebov, \enquote{Switching and sensing using kerr
  nonlinearity in quantum dots doped in metallic nanoshells,}
  {\protect\JournalTitle{The Journal of Physical Chemistry C}} \textbf{124},
  12065--12074 (2020).

\bibitem{lu2011ultrafast}
H.~Lu, X.~Liu, L.~Wang, Y.~Gong, and D.~Mao, \enquote{Ultrafast all-optical
  switching in nanoplasmonic waveguide with kerr nonlinear resonator,}
  {\protect\JournalTitle{Optics express}} \textbf{19}, 2910--2915 (2011).

\bibitem{ren2011nanostructured}
M.~Ren, B.~Jia, J.-Y. Ou, E.~Plum, J.~Zhang, K.~F. MacDonald, A.~E. Nikolaenko,
  J.~Xu, M.~Gu, and N.~I. Zheludev, \enquote{Nanostructured plasmonic medium
  for terahertz bandwidth all-optical switching,}
  {\protect\JournalTitle{Advanced Materials}} \textbf{23}, 5540--5544 (2011).

\bibitem{dani2009subpicosecond}
K.~M. Dani, Z.~Ku, P.~C. Upadhya, R.~P. Prasankumar, S.~Brueck, and A.~J.
  Taylor, \enquote{Subpicosecond optical switching with a negative index
  metamaterial,} {\protect\JournalTitle{Nano letters}} \textbf{9}, 3565--3569
  (2009).

\bibitem{cho2009ultrafast}
D.~J. Cho, W.~Wu, E.~Ponizovskaya, P.~Chaturvedi, A.~M. Bratkovsky, S.-Y. Wang,
  X.~Zhang, F.~Wang, and Y.~R. Shen, \enquote{Ultrafast modulation of optical
  metamaterials,} {\protect\JournalTitle{Optics express}} \textbf{17},
  17652--17657 (2009).

\bibitem{nikolaenko2010carbon}
A.~E. Nikolaenko, F.~De~Angelis, S.~A. Boden, N.~Papasimakis, P.~Ashburn,
  E.~Di~Fabrizio, and N.~I. Zheludev, \enquote{Carbon nanotubes in a photonic
  metamaterial: Giant ultrafast nonlinearity through plasmon-exciton coupling,}
  in \emph{Quantum Electronics and Laser Science Conference,}  (Optical Society
  of America, 2010), p. QTuD5.

\bibitem{lim2018ultrafast}
W.~X. Lim, M.~Manjappa, Y.~K. Srivastava, L.~Cong, A.~Kumar, K.~F. MacDonald,
  and R.~Singh, \enquote{Ultrafast all-optical switching of germanium-based
  flexible metaphotonic devices,} {\protect\JournalTitle{Advanced materials}}
  \textbf{30} (2018).

\bibitem{gholipour2013all}
B.~Gholipour, J.~Zhang, K.~F. MacDonald, D.~W. Hewak, and N.~I. Zheludev,
  \enquote{An all-optical, non-volatile, bidirectional, phase-change
  meta-switch,} {\protect\JournalTitle{Advanced materials}} \textbf{25},
  3050--3054 (2013).

\bibitem{bergamini2021single}
L.~Bergamini, B.~Chen, D.~Traviss, Y.~Wang, C.~H. de~Groot, J.~M. Gaskell,
  D.~W. Sheel, N.~Zabala, J.~Aizpurua, and O.~L. Muskens,
  \enquote{Single-nanoantenna driven nanoscale control of the vo2 insulator to
  metal transition,} {\protect\JournalTitle{Nanophotonics}}  (2021).

\bibitem{fang2015controlling}
X.~Fang, K.~F. MacDonald, and N.~I. Zheludev, \enquote{Controlling light with
  light using coherent metadevices: all-optical transistor, summator and
  invertor,} {\protect\JournalTitle{Light: Science \& Applications}}
  \textbf{4}, e292--e292 (2015).

\bibitem{plum2017controlling}
E.~Plum, K.~F. MacDonald, X.~Fang, D.~Faccio, and N.~I. Zheludev,
  \enquote{Controlling the optical response of 2d matter in standing waves,}
  {\protect\JournalTitle{ACS Photonics}} \textbf{4}, 3000--3011 (2017).

\bibitem{roger2015coherent}
T.~Roger, S.~Vezzoli, E.~Bolduc, J.~Valente, J.~J. Heitz, J.~Jeffers, C.~Soci,
  J.~Leach, C.~Couteau, N.~I. Zheludev \emph{et~al.}, \enquote{Coherent perfect
  absorption in deeply subwavelength films in the single-photon regime,}
  {\protect\JournalTitle{Nature communications}} \textbf{6}, 1--5 (2015).

\bibitem{fu2012all}
Y.~Fu, X.~Hu, C.~Lu, S.~Yue, H.~Yang, and Q.~Gong, \enquote{All-optical logic
  gates based on nanoscale plasmonic slot waveguides,}
  {\protect\JournalTitle{Nano letters}} \textbf{12}, 5784--5790 (2012).

\bibitem{zhang2012controlling}
J.~Zhang, K.~F. MacDonald, and N.~I. Zheludev, \enquote{Controlling
  light-with-light without nonlinearity,} {\protect\JournalTitle{Light: Science
  \& Applications}} \textbf{1}, e18--e18 (2012).

\bibitem{wan2011time}
W.~Wan, \enquote{Time-reversed lasing and interferometric control of absorption
  (vol 331, pg 889, 2011),} {\protect\JournalTitle{Science}} \textbf{334},
  1058--1058 (2011).

\bibitem{noh2012perfect}
H.~Noh, Y.~Chong, A.~D. Stone, and H.~Cao, \enquote{Perfect coupling of light
  to surface plasmons by coherent absorption,} {\protect\JournalTitle{Physical
  review letters}} \textbf{108}, 186805 (2012).

\bibitem{kang2014critical}
M.~Kang, Y.~Chong, H.-T. Wang, W.~Zhu, and M.~Premaratne, \enquote{Critical
  route for coherent perfect absorption in a fano resonance plasmonic system,}
  {\protect\JournalTitle{Applied Physics Letters}} \textbf{105}, 131103 (2014).

\bibitem{baranov2017coherent}
D.~G. Baranov, A.~Krasnok, T.~Shegai, A.~Al{\`u}, and Y.~Chong,
  \enquote{Coherent perfect absorbers: linear control of light with light,}
  {\protect\JournalTitle{Nature Reviews Materials}} \textbf{2}, 1--14 (2017).

\bibitem{kim2016general}
T.~Y. Kim, M.~A. Badsha, J.~Yoon, S.~Y. Lee, Y.~C. Jun, and C.~K. Hwangbo,
  \enquote{General strategy for broadband coherent perfect absorption and
  multi-wavelength all-optical switching based on epsilon-near-zero multilayer
  films,} {\protect\JournalTitle{Scientific reports}} \textbf{6}, 1--11 (2016).

\bibitem{ye2016coherent}
Y.~Ye, D.~Hay, and Z.~Shi, \enquote{Coherent perfect absorption in chiral
  metamaterials,} {\protect\JournalTitle{Optics letters}} \textbf{41},
  3359--3362 (2016).

\bibitem{zhang2008plasmon}
S.~Zhang, D.~A. Genov, Y.~Wang, M.~Liu, and X.~Zhang, \enquote{Plasmon-induced
  transparency in metamaterials,} {\protect\JournalTitle{Physical review
  letters}} \textbf{101}, 047401 (2008).

\bibitem{ling2018polarization}
Y.~Ling, L.~Huang, W.~Hong, T.~Liu, J.~Luan, W.~Liu, J.~Lai, and H.~Li,
  \enquote{Polarization-controlled dynamically switchable plasmon-induced
  transparency in plasmonic metamaterial,} {\protect\JournalTitle{Nanoscale}}
  \textbf{10}, 19517--19523 (2018).

\bibitem{scully1991enhancement}
M.~O. Scully, \enquote{Enhancement of the index of refraction via quantum
  coherence,} {\protect\JournalTitle{Physical Review Letters}} \textbf{67},
  1855 (1991).

\bibitem{panahpour2019refraction}
A.~Panahpour, A.~Mahmoodpoor, and A.~V. Lavrinenko, \enquote{Refraction
  enhancement in plasmonics by coherent control of plasmon resonances,}
  {\protect\JournalTitle{Physical Review B}} \textbf{100}, 075427 (2019).

\bibitem{Gunay2020continuously}
M.~Günay, Y.-L. Chuang, and M.~E. Tasgin, \enquote{Continuously-tunable
  cherenkov-radiation-based detectors via plasmon index control,}
  {\protect\JournalTitle{Nanophotonics}} \textbf{9}, 1479--1489 (2020).

\bibitem{Yuce2020Ultra}
E.~Yuce, A.~Kemal~Demir, Z.~Artvin, R.~Sahin, A.~Bek, and M.~E. Tasgin,
  \enquote{Ultra-large actively tunable photonic band gaps via plasmon-analog
  of index enhancement,} {\protect\JournalTitle{arXiv}} \textbf{07132}, 2
  (2020).

\bibitem{de2011dispersed}
A.~De~Luca, M.~P. Grzelczak, I.~Pastoriza-Santos, L.~M. Liz-Marzan, M.~La~Deda,
  M.~Striccoli, and G.~Strangi, \enquote{Dispersed and encapsulated gain medium
  in plasmonic nanoparticles: a multipronged approach to mitigate optical
  losses,} {\protect\JournalTitle{ACS nano}} \textbf{5}, 5823--5829 (2011).

\bibitem{infusino2014loss}
M.~Infusino, A.~De~Luca, A.~Veltri, C.~Vazquez~Vazquez, M.~A. Correa-Duarte,
  R.~Dhama, and G.~Strangi, \enquote{Loss-mitigated collective resonances in
  gain-assisted plasmonic mesocapsules,} {\protect\JournalTitle{ACS Photonics}}
  \textbf{1}, 371--376 (2014).

\bibitem{de2014double}
A.~De~Luca, R.~Dhama, A.~Rashed, C.~Coutant, S.~Ravaine, P.~Barois,
  M.~Infusino, and G.~Strangi, \enquote{Double strong exciton-plasmon coupling
  in gold nanoshells infiltrated with fluorophores,}
  {\protect\JournalTitle{Applied Physics Letters}} \textbf{104}, 103103 (2014).

\bibitem{dhama2016broadband}
R.~Dhama, A.~Rashed, V.~Caligiuri, M.~E. Kabbash, G.~Strangi, and A.~De~Luca,
  \enquote{Broadband optical transparency in plasmonic nanocomposite polymer
  films via exciton-plasmon energy transfer,} {\protect\JournalTitle{Optics
  express}} \textbf{24}, 14632--14641 (2016).

\bibitem{liu2009plasmonic}
N.~Liu, L.~Langguth, T.~Weiss, J.~K{\"a}stel, M.~Fleischhauer, T.~Pfau, and
  H.~Giessen, \enquote{Plasmonic analogue of electromagnetically induced
  transparency at the drude damping limit,} {\protect\JournalTitle{Nature
  materials}} \textbf{8}, 758--762 (2009).

\bibitem{viitanen2002analytical}
A.~Viitanen, I.~Hanninen, and S.~Tretyakov, \enquote{Analytical model for
  regular dense arrays of planar dipole scatterers,}
  {\protect\JournalTitle{Progress In Electromagnetics Research}} \textbf{38},
  97--110 (2002).

\bibitem{alu2011optical}
A.~Alu and N.~Engheta, \enquote{Optical wave interaction with two-dimensional
  arrays of plasmonic nanoparticles,} in \emph{Structured Surfaces as Optical
  Metamaterials,}  (Cambridge University Press Cambridge, UK, 2011), pp.
  58--93.

\bibitem{sasikala2018all}
V.~Sasikala and K.~Chitra, \enquote{All optical switching and associated
  technologies: a review,} {\protect\JournalTitle{Journal of Optics}}
  \textbf{47}, 307--317 (2018).

\bibitem{davis2014all}
T.~J. Davis, D.~E. G{\'o}mez, and F.~Eftekhari, \enquote{All-optical modulation
  and switching by a metamaterial of plasmonic circuits,}
  {\protect\JournalTitle{Optics letters}} \textbf{39}, 4938--4941 (2014).

\bibitem{de2012gain}
A.~De~Luca, M.~Ferrie, S.~Ravaine, M.~La~Deda, M.~Infusino, A.~R. Rashed,
  A.~Veltri, A.~Aradian, N.~Scaramuzza, and G.~Strangi, \enquote{Gain
  functionalized core--shell nanoparticles: the way to selectively compensate
  absorptive losses,} {\protect\JournalTitle{Journal of Materials Chemistry}}
  \textbf{22}, 8846--8852 (2012).

\end{thebibliography}






\end{document}